\documentstyle[twoside,fleqn,espcrc2]{article}

\newcommand{\AmS}{{\protect\the\textfont2
  A\kern-.1667em\lower.5ex\hbox{M}\kern-.125emS}}

\def\lsi{\raise0.3ex\hbox{$<$\kern-0.75em\raise-1.1ex\hbox{$\sim$}}}
\def\gsi{\raise0.3ex\hbox{$>$\kern-0.75em\raise-1.1ex\hbox{$\sim$}}}

\newcommand{\R}{{\kern+.25em\sf{R}\kern-.78em\sf{I} 
  \kern+.78em\kern-.25em}}
\newcommand{\C}{{\kern+.25em\sf{C}\kern-.50em\sf{I} \kern+.50em\kern-.25em}}

\title{Chern-Simons theory on the lattice}

\author{W. Bietenholz
\address{ Institut f\"{u}r Physik, Humboldt Universit\"{a}t zu Berlin,
Invalidenstr. 110, D-10115 Berlin, Germany \\
$^{{\rm b}}$ Dept. of Physics, Nagoya University, Nagoya 464-8601, Japan\\
$^{{\rm c}}$ Dipartimento di Fisica e Sezione I.N.F.N., Universit\'{a} di
Perugia, Via A. Pascoli, 06123 Perugia, Italy}, 
\ J. Nishimura $^{{\rm b}}$ and \ P. Sodano $^{{\rm c}}$
\thanks{Talk presented by W.B.\ at Lattice 2002; HU-EP-02/30.}
}

\begin{document}

\begin{abstract}

We present new proposals for the representation of a Chern-Simons
term on the lattice. In the first part, such a term is constructed
from the fermion determinant, and in the second part directly
from the Abelian gauge term. In both cases, the parity transformation
is modified on the lattice without affecting its continuum limit.

\vspace*{-5mm}

\end{abstract}

\maketitle

We consider the problem of representing the Chern-Simons (CS) term
on a 3d lattice. To this end, we modify the parity 
transformation on the lattice in a smooth way, similar to L\"{u}scher's
lattice modification of the chiral symmetry in even dimensions \cite{ML}.
In the first part we deal with 3d Ginsparg-Wilson (GW) fermions and
construct a CS term from their determinant. In the second
part we consider a pure Abelian CS action without introducing
matter fields.
The first and second part are based on Ref.\ \cite{BN} and
Ref.\ \cite{BS}, respectively.

\vspace*{-2mm}

\section{GW FERMIONS IN 3 DIMENSIONS}

\vspace*{-1mm}

In $d=4$ the GW relation \cite{GW} for a lattice Dirac operator $D^{(4)}$ is
well-known by now, $\{ D^{(4)}, \gamma_{5} \} = 
a D^{(4)} \gamma_{5} D^{(4)}$. In general one also assumes 
$\gamma_{5}$-Hermiticity, $ D^{(4)\, \dagger} = \gamma_{5} D^{(4)}
\gamma_{5}$, so that the GW relation can be written as
$D^{(4)} +  D^{(4)\, \dagger} =  a D^{(4)\, \dagger} D^{(4)}$.
This $\gamma_{5}$-independent form can be adapted in
$d=3$ (or other odd dimensions),
\begin{equation}  \label{GWR}
D^{(3)} +  D^{(3)\, \dagger} =  a D^{(3)\, \dagger} D^{(3)} \ .
\end{equation}
\vspace*{-0.5mm}
Such a lattice Dirac operator describes massless fermions in $d=3$.
In analogy to $\gamma_{5}$-Hermiticity, we assume in addition
the (natural) property
\begin{eqnarray}
D^{(3)\, \dagger}(U^{P}) &=& P D^{(3)\, \dagger}(U) P \ , 
\qquad {\rm where} \nonumber \\
\ U_{\mu}^{P}(x) &=& U_{\mu}(-x -a \hat \mu )^{\dagger} \ .
\end{eqnarray}
$P$ is the (standard) parity operator, and $U^{P}$ is the parity
transformed gauge configuration.

As in even dimensions, solutions to the GW relation (\ref{GWR})
take the form \cite{KN}
\begin{equation}  \label{VV}
D^{(3)} = \frac{1}{a} (1 - V) \ , \ V {\rm ~ unitary}
\end{equation}
where $V = 1 - a D_{cont} + O(a^{2})$.
The fermion action $S = a^{3} \sum_{x} \bar \psi (x) D^{(3)}(U) \psi (x)$
is exactly invariant under the lattice modified parity transformation
\begin{equation}  \nonumber
\psi (x)\to i P V \psi (x), \ \bar \psi (x) \to i \bar \psi (x) P, \
U_{\mu} \to U_{\mu}^{P}.
\end{equation}
The factor $V$ in the transformation of $\psi$
is a lattice modification in $O(a)$ (alternatively it could also
be attached to $\bar \psi$).

The measure transforms under the above transformation as
\begin{equation}  
d \bar \psi \, d \psi \to ({\rm det} V)^{-1} d \bar \psi \, d \psi
\end{equation}
which gives rise to the parity anomaly.
The situation is again similar to chiral symmetry in even
dimensions: a local, undoubled 3d fermion cannot be P invariant.
The Wilson term means a rough P breaking in the action, whereas
the GW fermion has a non-trivial transformation term of the modified
P symmetry in the measure; both types of symmetry breaking reproduce
the correct anomaly (see below).

Explicit GW operators $D^{(3)}$ are obtained by inserting
\begin{equation}
V = A / \sqrt{A^{\dagger}A} \ , \ A = 1 -a D_{0} \ ,
\end{equation}
into eq.\ (\ref{VV}), where $D_{0}$ is some lattice Dirac operator
(local, free of doubling), e.g.\ the Wilson operator $D^{(3)}_{w}$ 
\cite{KN}. Kikukawa and Neuberger further
observed the following phase relation
\cite{KN}
\begin{equation}  \label{phase}
2 \, {\rm arg} ({\rm det} D^{(3)}) = {\rm arg} ({\rm det} A) \ .
\end{equation}
Coste and L\"{u}scher studied the continuum limit of the
effective action ${\rm det} A$ \cite{CL}. In particular they considered
$A^{(n)} = (1 - a D^{(3)}_{w}) (1 - 2a D^{(3)}_{w})^{2n}$,
$n \in Z \!\!\! Z$. In a smooth gauge background, they found
\begin{equation}  \label{anomal}
^{lim}_{m,a \to 0} \ {\rm arg} ( {\rm det} A^{(n)}) =
(2n+1) \pi e^{2} S_{CS}
\end{equation}
where $m,e$ are the fermion mass and charge, and $S_{CS}$ is 
the CS action. We see that an infinite
set of universality classes (labeled by $n$) coexist for the
``P anomaly'' (which is therefore not an anomaly in the usual sense).

The standard overlap operator, which is obtained from
$A^{(0)} = 1 - D^{(3)}_{w}$, is in the class $n=0$, as eqs.\ 
(\ref{phase}, \ref{anomal}) show. However, it turns out that
the 3d GW fermions cover {\em all} universality classes, as we see
if we vary the terms $D_{0}$ and $R$ in the overlap solution
to the general GW relation \cite{WB}
\begin{eqnarray*}
D^{(3)} + D^{(3)\, \dagger} &=& 2a D^{(3)\, \dagger} R D^{(3)} \ , \\
D^{(3)} &=& \frac{1}{2a} \frac{1}{\sqrt{R}} (1 - A/\sqrt{A^{\dagger}
A} ) \frac{1}{\sqrt{R}} \\
A &=& 1 - 2a \sqrt{R} D_{0} \sqrt{R} \ ,
\end{eqnarray*}
where the term $R$ is local and not parity-odd. For example, we arrive
at the universality class $n$ for the choice $A=A^{(n)}$, $R_{x,y}
= \frac{1}{8n+2} \delta_{x,y}$, relying again on eqs.\ 
(\ref{phase}, \ref{anomal}).

These properties inspire the following ansatz for a {\em lattice}
CS term $S_{CS}^{lat}$ ,
\begin{equation}  \label{CSdet}
\exp (i S_{CS}^{lat}) = \frac{ {\rm det} A^{(0)} }
{ \vert {\rm det} A^{(0)} \vert }
\end{equation}
(where $A^{(0)}$ may also be replaced by another operator in the same
universality class). The phase is parity odd, the r.h.s.\ is manifestly
gauge invariant, and the normalization is taken so that
\begin{equation}
S_{SC}^{lat} \to S_{SC}^{lat} + 2 \pi \nu
\end{equation}
under a gauge transformation with winding number $\nu$.
Considering the above properties, we think that
this definition may capture the topology on the lattice
(this is somehow similar to the index in even dimensions).

\vspace*{-1mm}

\section{A NEW APPROACH TO PURE ABELIAN CS GAUGE THEORY}

\vspace*{-1mm}

In this part, we do not introduce any matter fields, but we are
only concerned with a suitable discretization of the CS Lagrangian
\begin{equation}
{\cal L}_{CS} = A_{\mu} \epsilon_{\mu \alpha \nu }
\partial_{\alpha} A_{\nu} \ ,
\end{equation}
which represents a topological field theory \cite{Witten}.
It is used in condensed matter physics and in polymer physics
as a low energy effective action \cite{Zee}.

For the naive lattice discretization the functional integral does not
exist (not even after gauge fixing) due to the doubling problem, 
which is typical for linear derivatives \cite{FM}. 
More generally, this problem
persists for any local, gauge invariant, cubic symmetric and
parity-odd action \cite{BDS}.

Our new approach to circumvent this problem
modifies this time the parity transform of the lattice gauge field.
A previous work along this line \cite{FL} succeeded in constructing 
a corresponding lattice action, but the modified P transform is 
unfortunately non-local. Here we insist on its locality, which gives
us confidence about a safe continuum limit.

For the lattice action at $a=1$ we write 
\begin{equation}
S[A] = \frac{1}{(2\pi )^{3}} \int_{-\pi}^{\pi} d^{3}p \, A_{\mu}(-p)
C_{\mu \nu}(p) A_{\nu}(p) \ .
\end{equation}
Now we follow the fermionic procedure: the perfect lattice action
for free fermions was derived analytically \cite{GW,perfect}.
If the RG blocking is done with a Gaussian of coefficient $R^{-1}$,
the perfect propagator obeys the GW relation with kernel $R$, which was
then extended to a general criterion for chirality.
To adapt this strategy to the current problem, we first constructed
a perfect lattice CS action, which is a straightforward application of 
the RG blocking of non-compact gauge fields \cite{perfect}.
The resulting perfect term $C_{\mu \nu}(p)$ has a property, which
is very similar to the GW relation; we denote it as the 
{\em Chern-Simons-Ginsparg-Wilson relation} (CSGWR)
\begin{equation}
C_{\mu \nu} + C_{\mu \nu}^{P} = C_{\mu \rho}^{P} C_{\rho \nu}
\end{equation}
(where we set the blocking term $R_{\mu \nu} = \frac{1}{2} 
\delta_{\mu \nu}$ for simplicity). We only use the perfect action 
to motivate this relation as a criterion for a smooth parity
breaking on the lattice. Now we move on to look for simpler solutions
to the CSGWR.

First, the overlap type of solution can also be applied
here \cite{BeS}: let $\Gamma_{\mu \nu} = C_{\mu \nu} - \delta_{\mu \nu}$
so that the CSGWR reads $\Gamma^{P}_{\mu \rho } \Gamma_{\rho \nu}
= \delta_{\mu \nu}$. Solutions are given by $\Gamma = \Gamma^{(0)}
[ \Gamma^{(0) \, P} \Gamma^{(0)}]^{-1/2}$. In analogy to the Wilson 
fermion, we can choose for instance the Fr\"{o}hlich-Marchetti action
\cite{FM} to fix $\Gamma^{(0)}$.

Now we look for even simpler, polynomial solutions to
the CSGWR, starting from the ansatz
\begin{eqnarray}  \label{Cform}
C_{\mu \nu}(p) &=& L_{\mu \nu} (p) v(p) + M_{\mu \nu}(p) w(p) \\
L_{\mu \nu}(p) &=& -i \epsilon_{\mu \alpha \nu} \hat p_{\alpha} \, ,
\ M_{\mu \nu}(p) = \hat p^{2} \delta_{\mu \nu} - \hat p_{\mu} 
\hat p_{\nu}   \nonumber
\end{eqnarray}
where $\hat p_{\alpha} = 2 \sin (p_{\alpha}/2)$. The functions
$v,\, w$ are even, local, lattice isotropic and $v(p) = 1 + O(p^{2})$.
The Maxwell term $M$ is parity even, so any $w \neq 0$ breaks the
anti-parity of $C$ in $O(a^{2})$ ($M$ is similar to the Wilson 
term for fermions). Fr\"{o}hlich and Marchetti set $v=w=1$ --- which 
implies a rough symmetry breaking --- but here we seek a smooth breaking,
which obeys the CSGWR.
Due to the identities $L^{P}_{\mu \rho} L_{\rho \nu}= - M_{\mu \nu}$
and $M^{P}_{\mu \rho} M_{\rho \nu}= \hat p^{2} M_{\mu \nu}$, the terms
reproduce themselves on the r.h.s.\ of the CSGWR. 
To solve it we have to require
\begin{equation}  \label{vw}
w(p) = \, [ \, 1 - \sqrt{1 + \hat p^{2} v^{2}(p)} \, ] \, / \hat p^{2} \ .
\end{equation}
For any local choice of $v$, the locality of $w$ is guaranteed as well.
These solutions are gauge invariant, as the identities
$\hat p_{\mu} L_{\mu \nu}(p) =0$ and $\hat p_{\mu} M_{\mu \nu}(p)=0$
reveal.

The question is now if we find a locally modified P transform
which keeps these solutions exactly invariant, i.e.
\begin{equation}  \label{modinv}
C_{\mu \nu} = - [ \delta_{\mu \rho} + X_{\mu \rho} ] C_{\rho \sigma}^{P}
[ \delta_{\sigma \nu} + X_{\sigma \nu} ] \ .
\end{equation}
For the modification term $X=O(a)$ we use the same ansatz
as for $C$,
\begin{equation}  \label{Xmod}
X_{\mu \nu}(p) = L_{\mu \nu} (p) x(p) + M_{\mu \nu}(p) y(p) \ ,
\end{equation}
where $x,y$ are even and lattice isotropic.

Inserting the ans\"{a}tze (\ref{Cform}) and (\ref{Xmod}) into
the condition (\ref{modinv}), and requiring the CSGWR to hold
(i.e.\ implementing eq.\ (\ref{vw})), 
we can express for instance $y,\, v, \, w$ in terms of $x$,
\begin{eqnarray}
y &=& [ \sqrt{1 + \hat p^{2} x^{2}} - 1 ] / \hat p^{2} \ , \nonumber \\
v &=& -2x \sqrt{1 + \hat p^{2} x^{2} } \ , \ \
w \, = \, -2x^{2} \ .
\end{eqnarray}
In particular,
any local choice for $x$ leads to local terms $v,w$ and $y$. 
To provide the right form of $v$ --- so that $C$ has the
correct continuum limit --- we need $x = -\frac{1}{2} + O(p^{2})$.
For instance, we can set $x=-1/2$, which
implies $y = [ \sqrt{1 + \hat p^{2}/4} -1 ] / \hat p^{2}$,
$v = \sqrt{1 + \hat p^{2}/4}$, $w=-1/2$.

Therefore, we found explicit, simple solutions to the CSGWR,
which are exactly odd under a lattice modified P transform.
Both, the action and the transformation term are {\em local}.
Therefore, we are on safe grounds to recover the correct continuum 
limit, which is a topological model.
However, it is an open question if our lattice formulation
is topological already.\\

\vspace*{-3mm}

To summarize, we note the CS action of the Part 1, $S_{CS}^{lat}$ in 
eq.\ (\ref{CSdet}), is not in conflict with 
the No-Go theorem of Ref.\ \cite{BDS}
because it is not bilinear in $A_{\mu}$, $A_{\nu}$.
Part 2 circumvents this theorem by modifying the parity
of $A_{\mu}$ on the lattice.

\vspace*{0.3mm}
\noindent
{\small\it We thank C.\ Diamantini, K.\ Jansen, F.\ Klinkhamer,
M.\ L\"{u}scher and U.-J.\ Wiese for useful comments.}

\end{document}